\documentclass[preprint,12pt]{elsarticle}
\usepackage{amssymb}

\journal{Physics Letters B}

\begin{document}

\makeatletter
\newdimen\saveparindent
\saveparindent=\parindent
\newdimen\saveparskip
\saveparskip=\parskip
\def\tablenotemark#1{\rlap{$^{\rm #1}$}}
\long\def\tablenotetext#1#2{\vtop{\vskip2pt
\uncentering\noindent\setbox0=\hbox{#1}\hskip\saveparindent\ifdim\wd0>1pt
$^{\rm #1}$ \fi{\ignorespaces #2}\vskip1sp}}
\def\uncentering{  \let\\\@normalcr
\rightskip0pt \leftskip0pt
\parindent\saveparskip \parfillskip0pt plus 1fil\relax}
\makeatother%

\begin{frontmatter}



\title{High-spin transition quadrupole moments in neutron-rich Mo and Ru nuclei:
testing $\gamma $ softness?}


\author{J.B. Snyder$^a$, W. Reviol$^b$, D.G. Sarantites$^b$, A.V.~Afanasjev$^c$,
R.V.F.~Janssens$^d$, H.~Abusara$^e$, M.P.~Carpenter$^d$, X.~Chen$^b$, C.J.~Chiara$^{d,f}$,
J.P.~Greene$^d$, T.~Lauritsen$^d$, E.A.~McCutchan$^{d,g}$,
D.~Seweryniak$^d$, S.~Zhu$^d$}

\address{$^a$ Physics Department, Washington University, St. Louis, MO 63130, USA}
\address{$^b$ Chemistry Department, Washington University, St. Louis, MO 63130, USA}
\address{$^c$ Department of Physics and Astronomy, Mississippi State University,
Starkville, MS 39762, USA}
\address{$^d$ Physics Division, Argonne National Laboratory, Argonne, IL 60439, USA}
\address{$^e$ Department of Physics, Faculty of Science, An-Najah National University,
Nablus, Palestine}
\address{$^f$ Department of Chemistry and Biochemistry, University of Maryland,
College Park, MD 20742, USA}
\address{$^g$ National Nuclear Data Center, Brookhaven National Laboratory,
Upton, NY 11973, USA}

\begin{abstract}
The transition quadrupole moments, $Q_{t}$, of rotational bands in
the neutron-rich, even-mass $^{102-108}$Mo and $^{108-112}$Ru nuclei
were measured in the 8 to 16~$\hbar $ spin range with the  
Doppler-shift attenuation method. The nuclei were populated as
fission fragments from $^{252}$Cf fission.
The detector setup consisted of the Gammasphere spectrometer and the
HERCULES fast-plastic array.
At moderate spin, the $Q_{t}$ moments are found to be reduced with
respect to the values near the ground states.
Attempts to describe the observations in mean-field-based models,
specifically cranked relativistic Hartree-Bogoliubov theory,
illustrate the challenge theory faces and the difficulty to infer
information on $\gamma $ softness and triaxiality from the data.
\end{abstract}

\begin{keyword}
rotational
bands \sep transition quadrupole moments \sep triaxial nuclear shape
\sep cranked relativistic Hartree-Bogoliubov theory
\end{%
keyword}

\end{frontmatter}

Evidence for the presence of a stable triaxial nuclear shape has thus far
mainly been inferred from spectroscopic data at high spin. Recent examples
are the observation of wobbling bands in the rare-earth region \cite{Jen-02}
and of chiral structures in the lanthanide region \cite{Fra-01}. Another,
earlier example of the presence of a triaxial shape is associated with the
smooth terminating bands in the tin region. These are understood as
corresponding to a gradual change, over a large spin range, from a prolate
shape through the triaxial plane toward an oblate, non-collective shape \cite%
{Afa-99}. Another testing ground for asymmetric shapes is thought to be the
Zr-Mo-Ru region of neutron-rich nuclei (neutron number $N\geq 60$). This is
due in part to the presence at low excitation energy of bands built on a $%
\gamma $-vibrational state, an observation \cite{Sha-94} suggesting that the
potential energy surface (PES) of these nuclei is soft or unstable with
respect to the deformation parameter $\gamma $. This parameter measures the
degree of triaxiality of a quadrupole nuclear shape, which is only symmetric
if $\gamma $ is a multiple of $60^{\circ }$ (e.g., prolate for $\gamma
=0^{\circ }$, $120^{\circ }$). However, there is at present considerable
uncertainty on this issue despite the increasing available information on
level schemes \cite{NNDC} and lifetimes \cite{Smi-12}. In fact, the data on
ground-state bands in the nuclei of interest appear consistent with the
rotation of a prolate or near-prolate shape accompanied by the usual
alignment of a pair of nucleons at medium spin.

It should be pointed out that the near-yrast structure of the nuclei of
interest cannot be interpreted in terms of either wobbling or chirality. The
known transitions linking excited- and ground-band levels are predominantly
of the $\Delta I=2$ and $\Delta I=0$ type, in contrast with the selective $%
\Delta I=1$ linking transitions associated with wobbling motion.
Furthermore, the alignment properties exhibited by the excited and the
ground-state bands differ significantly, while they are essentially
indistinguishable in the case of wobbling. The presence of chiral bands
indicates that the excited configuration, on which these bands are built, is
associated with a triaxial shape. It has been suggested that the off-yrast
structure of $^{110,112}$Ru may contain candidates for such bands \cite%
{Luo-PLB}. However, for $\gamma $-soft nuclei, the PES of an excited
configuration and of the vacuum configuration will be different, due to the
deformation driving properties of the orbitals in the former configuration
and to differences in pairing. Hence, the presence of chiral bands at
moderately high spin does not imply the onset of triaxiality near the ground
state.

In this letter, transition quadrupole moments, $Q_{t}$, in the even-mass
nuclei $^{102-108}$Mo and $^{108-112}$Ru are reported. The $Q_{t}$ values
are the result of lifetime measurements using the Doppler-shift attenuation
method (DSAM). The measurements cover a large number of transitions in the
respective ground-state bands, in the $6\leq I\leq 18$ spin range, and a
number of transitions in the $\gamma $ bands. The data sets are augmented by 
$Q_{t}$ values from the literature, including lower-spin transitions
accessible by the recoil-distance Doppler-shift method. The $Q_{t}$ values
and the moments of inertia are compared with the results of cranked
relativistic Hartree-Bogoliubov calculations with approximate
particle-number projection by means of the Lipkin-Nogami method (called
CRHB+LN hereafter) \cite{Afa-00}. The goal of the study is to examine
possible signatures for a shape change in the ground-state band as a
function of spin, herewith addressing the possibility of a rotation-induced
triaxial shape. The comparison between the data and the calculations also
addresses two related issues: the recent prediction of a stable triaxial
ground state \cite{Moe-08}, and the predicted competition between a prolate
and an oblate shape near the ground state \cite{Rodriguez-PLB} and in the
ground-state band \cite{Smi-12}.

As was the case in Ref. \cite{Smi-12}, neutron-rich nuclei in this mass
region were populated via spontaneous fission. A 230-$\mu $Ci $^{252}$Cf
source, mounted on a Pt backing of thickness 440 mg/cm$^{2}$, was used. The
source was covered by a 240-$\mu $g/cm$^{2}$ Au foil. The experiment ran for
18 days with a detector combination consisting of 98 Compton-suppressed Ge
spectrometers of Gammasphere \cite{Lee-90} and the HERCULES array. This
array of 64 fast-plastic detectors was designed as an evaporation-residue
counter for in-beam studies \cite{Rev-05}. In the present experiment,
HERCULES served a two-fold purpose: (i) It helped to determine the fission
axis, thereby providing an orientation axis for the emission angles of the $%
\gamma $ rays from fission fragments detected in Gammasphere. (ii) It
provided an efficient way to gate on either the light- or heavy-mass
fragment, based on the measured pulse height and time-of-flight with respect
to the $\gamma $-ray flash. A total of $2.1\cdot 10^{9}$ fragment-$\gamma
^{4}$ quintuple coincidence events were recorded.

\begin{figure}[th]
\includegraphics[scale=0.73]{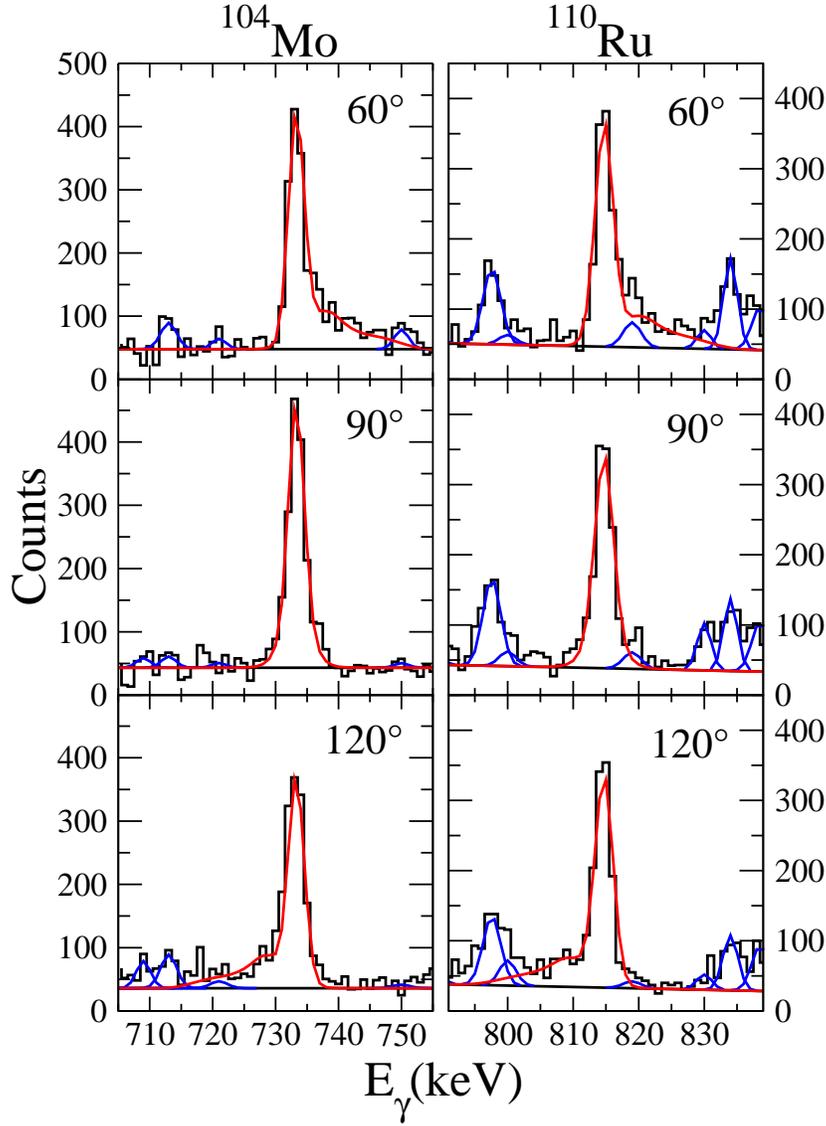}
\caption{Sample angle-sorted $\protect\gamma $-ray spectra ($\protect\xi %
=60^{\circ },90^{\circ },120^{\circ }$) and line-shape fits (red). Left: the
set for the 733.6-keV, $10_{1}^{+}\rightarrow 8_{1}^{+}$ transition in $%
^{104}$Mo. Right: the set for the 815.0-keV, $10_{1}^{+}\rightarrow
8_{1}^{+} $ transition in $^{110}$Ru. The spectra are the results of a
procedure with a fragment gate (see text) and a $\protect\gamma $-$\protect%
\gamma $ gate, where the gating transitions are below the analyzed
transition in the level scheme. Known contaminant lines (blue) are taken
into account in the fits.}
\label{one}
\end{figure}

Item (i) above is of importance as it enables DSAM lifetime measurements
with the usual approach; i.e., the analysis of asymmetric line shapes can be
performed, in contrast to the situation in Ref. \cite{Smi-12}. The fission
axis coincides with the line between the source position and the HERCULES
detector that is hit by a fragment. The velocity vector of the complimentary
fragment, slowing in the Pt backing and emitting the $\gamma $ ray of
interest, is collinear with this line (due to momentum conservation). The
emission angle of the $\gamma $\ ray of interest with respect to the fission
axis is denoted by $\xi $. Examples for angle-sorted $\gamma $-ray
coincidence spectra gated with the heavy-mass fragment are displayed in Fig. %
\ref{one}; more details are presented elsewhere \cite{Snyder}.

The line shapes for the various transitions are fitted with the code of Ref. 
\cite{Wel-96} and lifetimes are extracted. The light $^{252}$Cf fragments
have an average initial velocity of 0.046c. For the slowing of the ions in
the Pt backing, the stopping powers of both the SRIM package \cite{Zie-08}
and the Northcliffe-Schilling description \cite{Nor-70} were considered. The
two treatments differ in stopping powers by about 11\%. This difference,
along with uncertainties in the initial velocity and the transition energy,
were incorporated into the systematic uncertainty for each extracted
lifetime. The results are reported in Table \ref{Table-I}; they deal only
with stretched $E2$ transitions.

From the measured lifetimes, the reduced transition probabilities, $B(E2)$,
were derived. These were translated into $Q_{t}$ values according to the
expression 
\begin{equation}
B(E2;I_{i}\rightarrow I_{f})=\frac{5}{16\pi }\cdot \langle
I_{i}K_{i}(I_{i}-I_{f})(K_{i}-K_{f})\mid I_{f}K_{f}\rangle ^{2}\cdot
Q_{t}^{2},
\end{equation}

\noindent%
{}where the term in brackets is a Clebsch-Gordan coefficient determined by
the spin $I$ and the $K$ principal quantum number of the initial ($i$) and
final ($f$) state. Appropriate error propagation was taken into account. For
some of the transitions in a $\gamma $ band, the partial lifetimes according
to the intensity-branching ratios reported in Ref. \cite{NNDC} (or in a
footnote of Table \ref{Table-I}) were used. For all transitions in a $\gamma 
$ band, $K_{i}=$ $K_{f}=2$ was adopted. These values are confirmed \cite%
{Snyder} by ratios of measured $B(E2)$ values for different $\gamma $-band
to ground-state-band transitions from the same initial state according to
the Alaga rules \cite{Casten}.

\begin{figure}[th]
\begin{center}
\includegraphics[scale=0.54]{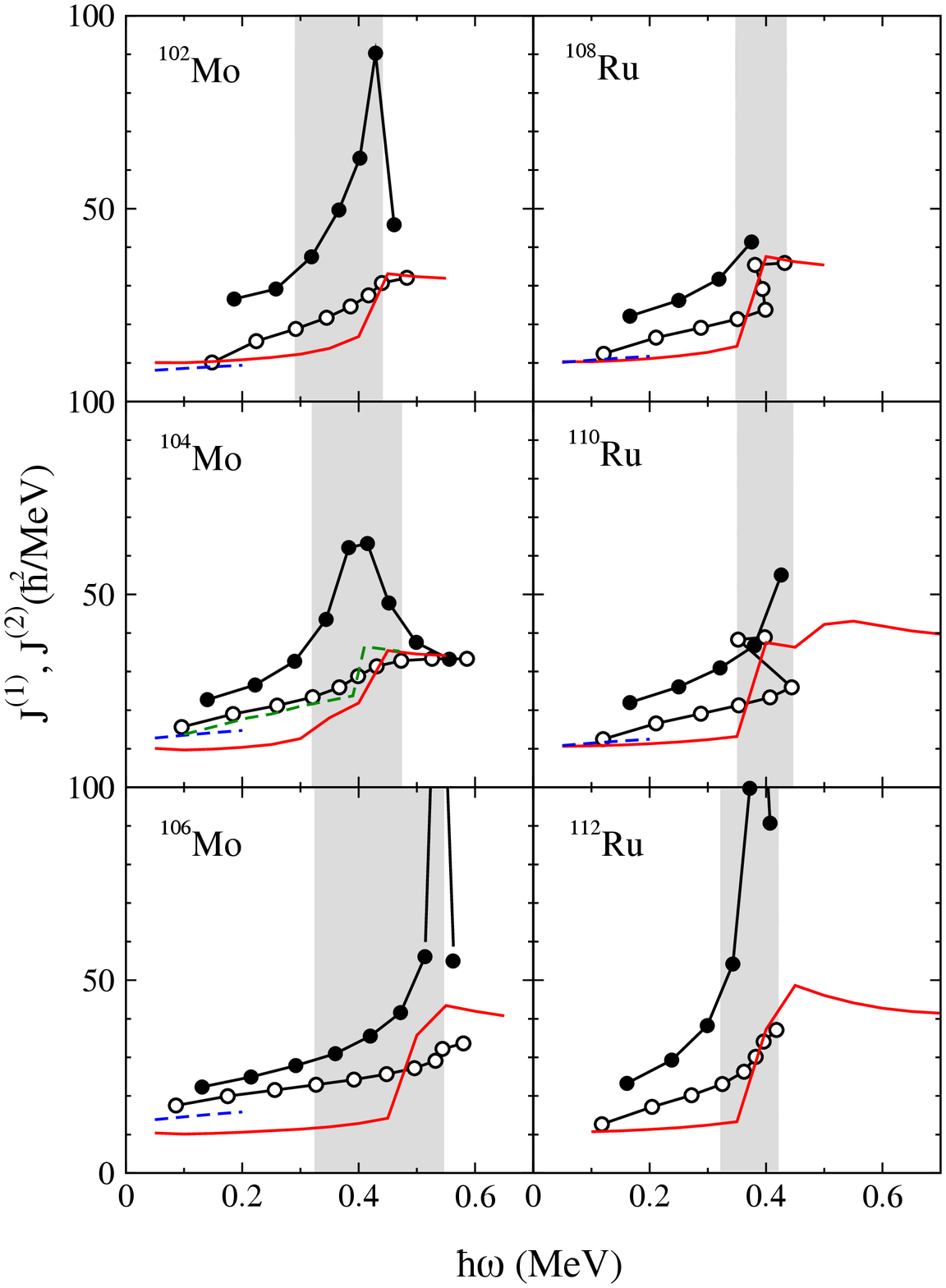}
\end{center}
\caption{Kinematic ($\Im ^{(1)}$) and dynamic ($\Im ^{(2)}$) moments of
inertia, as a function of rotational frequency ($\protect\omega $), for the
ground-state bands in $^{102-106}$Mo and $^{108-112}$Ru, based on the level
schemes in Ref. \protect\cite{NNDC}. The highest values of the $^{106}$Mo
and $^{112}$Ru $\Im ^{(2)}$ moments are off-scale. The $^{108,110}$Ru $\Im
^{(2)}$ moments are truncated where the $\Im ^{(1)}$ moments show backbends.
Shaded areas represent the spin ranges of the $Q_{t}$ values in Tab. I. The
solid, red and dashed, green curves present the $\Im ^{(1)}$ moments for
near-oblate (triaxial) and near-prolate minima, respectively, from CRHB+LN
calculations. The blue, dashed lines display the $\Im ^{(1)}$ moments of
excited prolate minima at $\protect\omega \approx 0$, which become unstable
at higher frequencies. These lines are stretched for visibility.}
\label{two}
\end{figure}

The present discussion focusses on the ground-state bands and starts with
their alignment features. These are depicted in Fig. \ref{two} in terms of
the kinematic and dynamic moments of inertia as a function of rotational
frequency. The $^{106}$Mo and $^{108}$Mo nuclei have essentially the same
characteristics and the latter case is omitted for brevity. As indicated by
the shaded area in each plot, the spin range where $Q_{t}$ values are now
available, overlaps in part with the band-crossing region. In $^{104}$Mo,
for example, this region is centered around $\hbar \omega =0.4$\ MeV and $%
I=12$. In all nuclei under discussion, the rise of the moments of inertia is
attributed to the rotational alignment of a pair of $h_{11/2}$ neutrons \cite%
{Ska-97}.

\begin{figure}[tph]
\begin{center}
\includegraphics[scale=1.10]{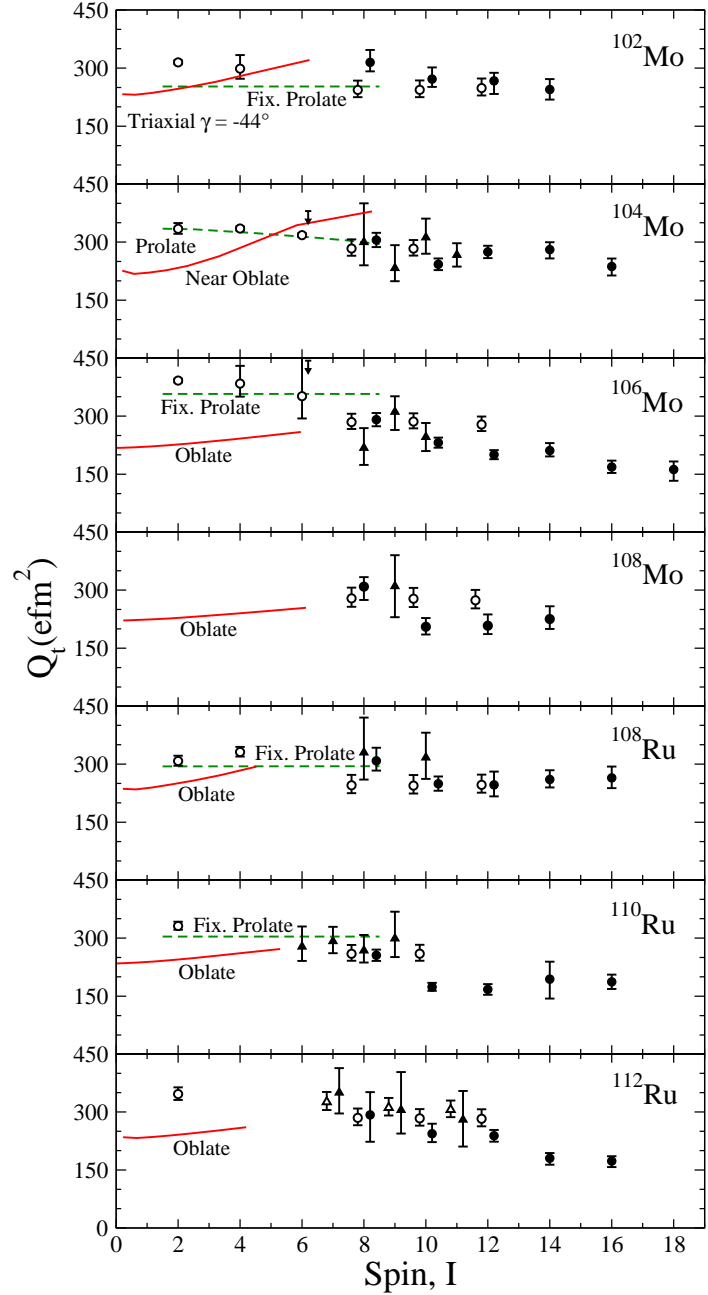}
\end{center}
\caption{Transition quadrupole moment as a function of spin for ground-state
bands (circles) and $\protect\gamma $ bands (triangles up) in $^{102-108}$Mo
and $^{108-112}$Ru. Data from this work are given as full symbols, data
reported in the literature as open ones. The CRHB+LN results are drawn with
the convention of Fig.~\protect\ref{two} and are labeled accordingly. }
\label{three}
\end{figure}

The $Q_{t}$ values, as a function of spin, for the in-band transitions in $%
^{102-108}$Mo and $^{108-112}$Ru are provided in Fig. \ref{three} with full
symbols. They are combined with previous results from the literature \cite%
{NNDC,Smi-12}, shown as open symbols. A distinction is made between the $%
Q_{t}$ values for the ground-state and $\gamma $ bands by circles and
triangles, respectively. The values for the two types of bands are
comparable in magnitude.

For the ground-state bands the following observations can be made: The $%
Q_{t} $ values decrease with increasing spin and this behavior is
accentuated in the heavier isotopes. This observation holds for both Mo ($%
^{102,104}$Mo vs. $^{106,108}$Mo) and Ru ($^{108}$Ru vs. $^{110,112}$Ru)
isotopes: the weighted-average values for $8\leq I\leq 16$ of $192\pm 16$ efm%
$^{2}$ and $198\pm 26$ efm$^{2}$ for $^{110,112}$Ru are to be compared with $%
261\pm 11$ efm$^{2}$ for $^{108}$Ru. Moreover, these averages for $%
^{110,112} $Ru are also somewhat smaller than those for $^{106,108}$Mo.
Hence, the decrease in the $Q_{t}$ values seems more severe in the Ru
isotopes than in the available Mo nuclei. This observation suggests a
dependence of $Q_{t}(I)$ on $Z$ and $N$. It is worth noting that a reduction
of $Q_{t}$ moments with increasing spin is also seen in $^{74}$Kr \cite%
{Valiente-Dobon} and in the rare-earth region \cite{Joh-92}. In these
nuclei, this reduction has been interpreted as being due to a $\gamma $ soft
PES polarized by rotation-aligned quasiparticles inducing a triaxial shape.

As stated above, the Mo and Ru nuclei under investigation are thought to be
characterized by $\gamma $-soft energy surfaces and their successful
description would be expected to use mean-field based models. Unfortunately,
as discussed in Sec. 4.1. of Ref. \cite{Moe-08}, no consistent picture
emerges from the model calculations and different methods reach different
conclusions. These calculations face two principal difficulties. First,
strong shape variations with $Z$ and $N$ are expected that can be attributed
to shell effects in the single-particle spectrum. Hence, the results depend
sensitively on the adopted single-particle energies, the accuracy of which
is model and parameter dependent \cite{Afa-11,Rodriguez-PRC}. Second, the
results of calculations strongly depend on the treatment of pairing as
exemplified below.

In this letter, the covariant density functional theory of Ref. \cite{Afa-00}
is applied. The first task was to perform PES calculations as a function of
the $\gamma $ degree of freedom and the quadrupole-deformation parameter $%
\beta _{2}$. This was done for selected Mo isotopes with a code for triaxial
relativistic mean-field theory plus BCS with separable pairing (called
RMF+BCS hereafter). The second task was to calculate the alignment and
deformation properties of the bands and this was done within the CRHB+LN
approach \cite{Afa-00}, where the Gogny D1S force was used in the pairing
channel. This code does not have constraints on the diagonal and
off-diagonal elements of the quadrupole-moment tensors, $Q_{22}$ and $Q_{20}$%
. As a result, the solution is restricted to local minima; i.e., the
observables of interest, $Q_{t}$ and $\Im ^{(1)}$, are calculated at the
equilibrium deformations of these minima, which change with frequency. In
both calculations the NL3* parametrization of the RMF Lagrangian \cite%
{Lal-09} was used.

\begin{figure}[tph]
\begin{center}
\includegraphics[scale=0.51]{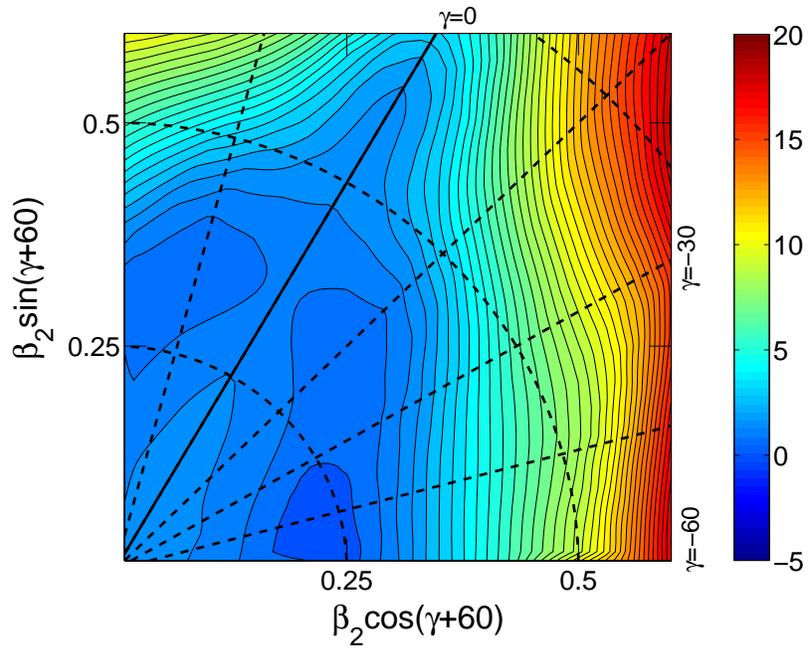}
\end{center}
\caption{PES for $^{104}$Mo from a triaxial RMF+BCS calculation. The energy
difference between equipotential curves is 0.5 MeV. The color scale shown at
the right has the unit MeV.}
\label{PES}
\end{figure}

A sample result of the triaxial RMF+BCS calculations with NL3* is shown in
Fig. \ref{PES}. The PES plot illustrates the $\gamma $\ softness in this
case. An oblate minimum and a shallow excited prolate minimum are seen. In
contrast, axial RMF+BCS calculations \cite{Lal-99} with the NL3
parametrization of the Lagrangian predict the prolate minimum to be the
lowest in $^{102,104}$Mo, whereas the oblate one becomes the lowest in the
heavier Mo isotopes. However, CRHB+LN calculations with the NL3
parametrization suggest that the oblate minimum is lowest by 20 and 240 keV
in $^{102,104}$Mo, respectively. These examples indicate that the energy
surfaces strongly depend on the treatment of pairing.

The results of the CRHB+LN calculations are first compared with the
experimental $\Im ^{(1)}$ moments in Fig. \ref{two}. The ground states of $%
^{102,104,106}$Mo are calculated to be triaxial $(\gamma \sim -44^{\circ })$%
, near-oblate $(\gamma \sim -53^{\circ })$, and oblate, respectively. These
solutions are energetically favored in the calculations. However, they fail
to reproduce the rise of the $\Im ^{(1)}$ moments with frequency. In Fig. %
\ref{three}, the triaxial and oblate solutions are represented by a full
curve. Substantial shape changes take place in the associated configurations
with increasing spin: the $\beta _{2}$ deformation increases while $\gamma $
drifts towards $-30^{\circ }$. The latter feature is pronounced in $^{104}$%
Mo, where $\gamma \thicksim $ $-30^{\circ }$ is reached at $I\sim 4$ and the 
$Q_{t}$ value rises accordingly. This prediction is in conflict with the
data, including the new $Q_{t}$ values, which show the opposite trend. A
similar situation occurs in the Ru isotopes. Thus, the interpretation of the
alignment and deformation properties of the observed bands in terms of
collective motion associated with oblate and near-oblate shapes faces
substantial difficulties. It turns out that it is also impossible to
describe the radii of neutron-rich Mo nuclei with such shapes \cite%
{Rodriguez-PLB}.

The alternative is to associate the observed bands with a prolate minimum,
although it is an excited one in Fig. \ref{PES}. The CRHB+LN calculations
indicate that, without constraining the $Q_{20}$ and $Q_{22}$ moments, the
solution in the local prolate minimum becomes unstable when $\omega $
increases. Only in the case of $^{104}$Mo is a solution obtained over a
significant frequency range. Figures\ \ref{two} and \ref{three} indicate
that the CRHB+LN prolate solution in $^{104}$Mo, represented by dashed,
green curves, provides a good description of the $\Im ^{(1)}$ moment and the
band-crossing frequency as well as the $Q_{t}$ values for $I\leq 8$. The
downslope of $Q_{t}$ with increasing $I$ is reproduced and is attributed to
a combined decrease in $\beta _{2}$ and increase in $\gamma $ deformation
induced by rotation. For the other nuclei, the prolate solution is only
stable at the lowest $\omega $ values. The $\Im ^{(1)}$ and $Q_{t}$ values
extracted from these minima agree rather well with experiment, though in a
limited range of frequency and spin. Due to this limitation, the trends in
the predicted low-spin and measured high-spin $Q_{t}$ values are, in the
case of $^{106}$Mo and $^{110}$Ru, not comparable. They can, however, be
viewed as complementary. It is worth noting that the current $Q_{t}$ data
are covering the band-crossing region; this spin range disappears in the
cranking calculations performed as a function of $\omega $ if the calculated
crossing is sharp \cite{Szymanski}.

The trend seen in the data appears to be consistent with rotation associated
with a near-prolate shape below the band crossing. Above it, a significant
excursion into the triaxial sector may be present according to the current
CRHB+LN calculations. These data represent a challenge for theoretical
calculations based on mean-field models. This is demonstrated here within
the CRHB+LN framework. The predicted oblate shape and a low-spin triaxial
shape with $\gamma \sim -44^{\circ }$ are ruled out by the data. For some of
the nuclei viz. $^{104}$Mo, the calculations with the prolate minimum
reproduce the observations, but reliable predictions with such a minimum
cannot be made for all the nuclei discussed. For nuclei with a very soft
PES, a description on the mean-field level may not be adequate and methods
beyond mean field may be required \cite{Ben-06,Li-09}, as correlations due
to configuration mixing and angular-momentum projection can affect the
relative energies of the various minima. However, such methods also depend
sensitively on the underlying single-particle structure which remains model
and parameter dependent \cite{Ben-06}. Thus, description within such a
framework will not necessarily provide reliable results. In addition, the
description of rotational spectra requires the use of a phenomenological
scaling factor for the moments of inertia, as time-odd mean fields are
neglected in the current realizations of these methods \cite{Li-09}.

In conclusion, $Q_{t}$ moments up to $I=16-18$ have been obtained for the $%
^{102-108}$Mo and $^{108-112}$Ru even, neutron-rich isotopes. A systematic
decrease of the $Q_{t}$ moments with spin is observed. The available data
remain a challenge for theory to explain.

\section*{Acknowledgments}

The authors thank J. Elson (WU) and J. Rohrer (ANL) for technical support.
The $^{252}$Cf source was provided by the Oak Ridge National Laboratory.
This work was supported by the US Department of Energy, Office of Nuclear
Physics, Grant Nos. DE-FG02-88ER40406, DE-FG02-07ER41459, DE-FG02-94ER40834,
and Contract No. DE-AC02-06CH11357.





\bibliographystyle{elsarticle-num}
\bibliography{<your-bib-database>}



\bigskip

\clearpage

\begin{table}[h]\centering%
\caption{Properties of transitions in $^{102-108}$Mo and  $^{108-112}$Ru for which
$Q_{t}$ values are obtained. Spin-parity assignments and transition energies are
adopted from Ref. \protect\cite{NNDC}.
Intensities are from the present work and are given relative to I$_{\gamma }$ $\equiv $100
for the corresponding $8_{1}^{+}\rightarrow 6_{1}^{+}$ transition. 
Lifetime and $Q_{t}$-value uncertainties contain both statistical and systematic errors.}
\label{Table-I}{\vspace { 3mm}}%

\begin{tabular}{ccccc}
\hline
$I_{i}^{\pi }\rightarrow I_{f}^{\pi }$ \tablenotemark{1}\  & E$_{\gamma }$
(keV) & I$_{\gamma }$\ \tablenotemark{2} & $\tau $ (ps) & $Q_{t}$ (efm$^{2}$)
\\ \hline
\multicolumn{5}{c}{$^{102}$Mo} \\ 
\multicolumn{1}{l}{$8_{1}^{+}\rightarrow 6_{1}^{+}$} & 690.9 & $\equiv $100
& 1.59$_{-0.32}^{+0.23}$ & 315$_{-23}^{+32}$ \\ 
\multicolumn{1}{l}{$10_{1}^{+}\rightarrow 8_{1}^{+}$} & 771.5 & 48.1 & 1.20$%
_{-0.27}^{+0.18}$ & 272$_{-20}^{+30}$ \\ 
\multicolumn{1}{l}{$12_{1}^{+}\rightarrow 10_{1}^{+}$} & 834.9 & 20.6 & 0.82$%
_{-0.13}^{+0.21}$ & 267$_{-34}^{+21}$ \\ 
\multicolumn{1}{l}{$14_{1}^{+}\rightarrow 12_{1}^{+}$} & 879.2 & 2.3 & 0.74$%
_{-0.16}^{+0.16}$ & 245$_{-26}^{+27}$ \\ 
\multicolumn{5}{c}{$^{104}$Mo} \\ 
\multicolumn{1}{l}{$6_{1}^{+}\rightarrow 4_{1}^{+}$} & 519.2 & 184 & $\geq $%
4.78\tablenotemark{3} & $\leq $380 \\ 
\multicolumn{1}{l}{$8_{1}^{+}\rightarrow 6_{1}^{+}$} & 641.7 & $\equiv $100
& 2.44$_{-0.30}^{+0.29}$ & 305$_{-18}^{+19}$ \\ 
\multicolumn{1}{l}{$8_{2}^{+}\rightarrow 6_{2}^{+}$} & 601.7 & 24.9 & 3.4$%
_{-1.4}^{+1.5}$ & 300$_{-60}^{+100}$ \\ 
\multicolumn{1}{l}{$9_{1}^{+}\rightarrow 7_{1}^{+}$} & 646.5 & 14.0 & 2.24$%
_{-0.73}^{+0.64}$ & 233$_{-34}^{+59}$ \\ 
\multicolumn{1}{l}{$10_{1}^{+}\rightarrow 8_{1}^{+}$} & 733.6 & 40.8 & 1.93$%
_{-0.24}^{+0.24}$ & 243$_{-15}^{+15}$ \\ 
\multicolumn{1}{l}{$10_{2}^{+}\rightarrow 8_{2}^{+}$} & 678.4 & 6.2 & 1.88$%
_{-0.59}^{+0.51}$ & 312$_{-42}^{+49}$ \\ 
\multicolumn{1}{l}{$11^{+}\rightarrow 9_{1}^{+}$} & 712.9 & 5.1 & 1.96$%
_{-0.44}^{+0.44}$ & 267$_{-30}^{+30}$ \\ 
\multicolumn{1}{l}{$12_{1}^{+}\rightarrow 10_{1}^{+}$} & 798.0 & 19.7 & 0.96$%
_{-0.11}^{+0.11}$ & 275$_{-16}^{+16}$ \\ 
\multicolumn{1}{l}{$14_{1}^{+}\rightarrow 12_{1}^{+}$} & 861.3 & 10.0 & 0.63$%
_{-0.09}^{+0.10}$ & 281$_{-23}^{+19}$ \\ 
\multicolumn{1}{l}{$16_{1}^{+}\rightarrow 14_{1}^{+}$} & 945.0 & 2.6 & 0.55$%
_{-0.10}^{+0.11}$ & 237$_{-23}^{+21}$ \\ 
\multicolumn{5}{c}{$^{106}$Mo} \\ 
\multicolumn{1}{l}{$4_{2}^{+}\rightarrow 2_{1}^{+}$} & 896.2 & 55.9 & 2.33$%
_{-0.29}^{+0.27}$ & 115$_{-17}^{+19}$\tablenotemark{4} \tablenotemark{5} \\ 
\multicolumn{1}{l}{$6_{1}^{+}\rightarrow 4_{1}^{+}$} & 511.2 & 252 & $\geq $%
3.80\tablenotemark{3} & $\leq $443 \\ 
\multicolumn{1}{l}{$8_{1}^{+}\rightarrow 6_{1}^{+}$} & 654.9 & $\equiv $100
& 2.43$_{-0.29}^{+0.29}$ & 291$_{-17}^{+17}$ \\ 
\multicolumn{1}{l}{$8_{2}^{+}\rightarrow 6_{2}^{+}$} & 631.0 & 12.1 & 3.46$%
_{-0.69}^{+0.71}$ & 218$_{-44}^{+51}$\tablenotemark{6} \\ 
\multicolumn{1}{l}{$9_{1}^{+}\rightarrow 7_{1}^{+}$} & 690.9 & 9.1 & 1.81$%
_{-0.47}^{+0.54}$ & 311$_{-47}^{+40}$ \\ 
\multicolumn{1}{l}{$10_{1}^{+}\rightarrow 8_{1}^{+}$} & 784.1 & 42.9 & 1.52$%
_{-0.17}^{+0.17}$ & 231$_{-13}^{+13}$ \\ 
\multicolumn{1}{l}{$10_{2}^{+}\rightarrow 8_{2}^{+}$} & 756.4 & 3.6 & 1.77$%
_{-0.52}^{+0.52}$ & 246$_{-36}^{+36}$ \\ 
\multicolumn{1}{l}{$12_{1}^{+}\rightarrow 10_{1}^{+}$} & 896.7 & 17.9 & 1.02$%
_{-0.12}^{+0.12}$ & 200$_{-12}^{+12}$ \\ 
\multicolumn{1}{l}{$14_{1}^{+}\rightarrow 12_{1}^{+}$} & 992.9 & 8.9 & 0.54$%
_{-0.10}^{+0.08}$ & 211$_{-15}^{+19}$ \\ 
\multicolumn{1}{l}{$16_{1}^{+}\rightarrow 14_{1}^{+}$} & 1051.5 & 5.2 & 0.63$%
_{-0.12}^{+0.12}$ & 169$_{-16}^{+16}$ \\ 
\multicolumn{1}{l}{$18_{1}^{+}\rightarrow 16_{1}^{+}$} & 1087.6 & 1.9 & 0.57$%
_{-0.15}^{+0.21}$ & 162$_{-29}^{+21}$ \\ \hline
\end{tabular}

\end{table}%

\addtocounter{table}{-1}
\begin{table}[h]\centering%
\caption{\textit{(Continued.)}}
\label{Table-I}{\vspace { 3mm}}%

\begin{tabular}{ccccc}
\hline
$I_{i}^{\pi }\rightarrow I_{f}^{\pi }$ \tablenotemark{1}\  & E$_{\gamma }$
(keV) & I$_{\gamma }$\ \tablenotemark{2} & $\tau $ (ps) & Q$_{t}$ (efm$^{2}$)
\\ \hline
\multicolumn{5}{c}{$^{108}$Mo} \\ 
\multicolumn{1}{l}{$8_{1}^{+}\rightarrow 6_{1}^{+}$} & 662.1 & $\equiv $100
& 2.05$_{-0.32}^{+0.46}$ & 309$_{-35}^{+24}$ \\ 
\multicolumn{1}{l}{$9_{1}^{+}\rightarrow 7_{1}^{+}$} & 707.0 & 25.0 & 1.6$%
_{-0.6}^{+1.2}$ & 310$_{-80}^{+80}$ \\ 
\multicolumn{1}{l}{$10_{1}^{+}\rightarrow 8_{1}^{+}$} & 776.6 & 57.7 & 2.02$%
_{-0.44}^{+0.39}$ & 205$_{-20}^{+22}$ \\ 
\multicolumn{1}{l}{$12_{1}^{+}\rightarrow 10_{1}^{+}$} & 872.0 & 19.6 & 1.08$%
_{-0.30}^{+0.22}$ & 208$_{-22}^{+29}$ \\ 
\multicolumn{1}{l}{$14^{+}\rightarrow 12_{1}^{+}$} & 945.6 & 8.2 & 0.61$%
_{-0.18}^{+0.14}$ & 225$_{-26}^{+33}$ \\ 
\multicolumn{5}{c}{$^{108}$Ru} \\ 
\multicolumn{1}{l}{$4_{2}^{+}\rightarrow 2_{1}^{+}$} & 940.5 & 32.2 & 1.87$%
_{-0.30}^{+0.30}$ & 78$_{-26}^{+26}$\tablenotemark{5} \\ 
\multicolumn{1}{l}{$8_{1}^{+}\rightarrow 6_{1}^{+}$} & 701.6 & $\equiv $100
& 1.53$_{-0.34}^{+0.25}$ & 308$_{-25}^{+34}$ \\ 
\multicolumn{1}{l}{$8_{2}^{+}\rightarrow 6_{2}^{+}$} & 657.8 & 20.8 & 2.1$%
_{-0.8}^{+1.3}$ & 330$_{-70}^{+90}$ \\ 
\multicolumn{1}{l}{$10_{1}^{+}\rightarrow 8_{1}^{+}$} & 798.3 & 41.6 & 1.20$%
_{-0.18}^{+0.17}$ & 249$_{-18}^{+19}$ \\ 
\multicolumn{1}{l}{$10_{2}^{+}\rightarrow 8_{2}^{+}$} & 730.0 & 11.5 & 1.25$%
_{-0.50}^{+0.44}$ & 317$_{-55}^{+64}$ \\ 
\multicolumn{1}{l}{$12_{1}^{+}\rightarrow 10_{1}^{+}$} & 788.1 & 19.9 & 1.28$%
_{-0.36}^{+0.31}$ & 246$_{-30}^{+34}$ \\ 
\multicolumn{1}{l}{$14_{1}^{+}\rightarrow 12_{1}^{+}$} & 762.2 & 5.1 & 1.34$%
_{-0.25}^{+0.21}$ & 260$_{-21}^{+24}$ \\ 
\multicolumn{1}{l}{$16_{1}^{+}\rightarrow 14_{1}^{+}$} & 863.6 & 2.3 & 0.69$%
_{-0.15}^{+0.14}$ & 264$_{-27}^{+29}$ \\ 
\multicolumn{5}{c}{$^{110}$Ru} \\ 
\multicolumn{1}{l}{$6_{2}^{+}\rightarrow 4_{2}^{+}$} & 599.8 & 62.5 & 4.12$%
_{-0.99}^{+0.94}$ & 278$_{-37}^{+52}$ \\ 
\multicolumn{1}{l}{$7_{1}^{+}\rightarrow 5_{1}^{+}$} & 645.5 & 20.8 & 3.03$%
_{-0.49}^{+0.50}$ & 292$_{-31}^{+37}$ \\ 
\multicolumn{1}{l}{$8_{1}^{+}\rightarrow 6_{1}^{+}$} & 705.3 & $\equiv $100
& 2.17$_{-0.25}^{+0.25}$ & 256$_{-15}^{+15}$ \\ 
\multicolumn{1}{l}{$8_{2}^{+}\rightarrow 6_{2}^{+}$} & 712.7 & 25.0 & 1.94$%
_{-0.38}^{+0.36}$ & 268$_{-31}^{+40}$ \\ 
\multicolumn{1}{l}{$9^{+}\rightarrow 7^{+}$} & 756.0 & 8.6 & 1.24$%
_{-0.57}^{+0.40}$ & 299$_{-48}^{+69}$ \\ 
\multicolumn{1}{l}{$10_{1}^{+}\rightarrow 8_{1}^{+}$} & 815.0 & 37.6 & 2.22$%
_{-0.27}^{+0.26}$ & 174$_{-10}^{+11}$ \\ 
\multicolumn{1}{l}{$12_{1}^{+}\rightarrow 10_{1}^{+}$} & 887.6 & 12.5 & 1.54$%
_{-0.25}^{+0.25}$ & 168$_{-14}^{+14}$ \\ 
\multicolumn{1}{l}{$14_{1}^{+}\rightarrow 12_{1}^{+}$} & 703.9 & 7.9 & 3.6$%
_{-1.2}^{+2.9}$ & 190$_{-50}^{+40}$ \\ 
\multicolumn{1}{l}{$16_{1}^{+}\rightarrow 14_{1}^{+}$} & 799.7 & 5.4 & 2.03$%
_{-0.40}^{+0.40}$ & 187$_{-19}^{+19}$ \\ 
\multicolumn{5}{c}{$^{112}$Ru} \\ 
\multicolumn{1}{l}{$7_{1}^{+}\rightarrow 5_{1}^{+}$} & 605.4 & 44.5 & 3.10$%
_{-1.12}^{+0.96}$ & 350$_{-54}^{+63}$ \\ 
\multicolumn{1}{l}{$8_{1}^{+}\rightarrow 6_{1}^{+}$} & 649.5 & $\equiv $100
& 2.5$_{-0.7}^{+1.8}$ & 290$_{-70}^{+60}$ \\ 
\multicolumn{1}{l}{$9_{1}^{+}\rightarrow 7_{1}^{+}$} & 693.3 & 26.5 & 1.84$%
_{-0.79}^{+1.01}$ & 305$_{-61}^{+98}$ \\ 
\multicolumn{1}{l}{$10_{1}^{+}\rightarrow 8_{1}^{+}$} & 723.3 & 55.9 & 2.06$%
_{-0.44}^{+0.37}$ & 244$_{-22}^{+26}$ \\ 
\multicolumn{1}{l}{$11_{1}^{+}\rightarrow 9_{1}^{+}$\tablenotemark{7}} & 
756.0 & 10.2 & 1.32$_{-0.70}^{+0.65}$ & 280$_{-69}^{+74}$ \\ \hline
\end{tabular}

\end{table}%

\addtocounter{table}{-1}
\begin{table}[h]\centering%
\caption{\textit{(Continued.)}}
\label{Table-I}{\vspace { 3mm}}%

\begin{tabular}{ccccc}
\hline
$I_{i}^{\pi }\rightarrow I_{f}^{\pi }$ \tablenotemark{1}\  & E$_{\gamma }$
(keV) & I$_{\gamma }$\ \tablenotemark{2} & $\tau $ (ps) & Q$_{t}$ (efm$^{2}$)
\\ \hline
\multicolumn{1}{l}{$12_{1}^{+}\rightarrow 10_{1}^{+}$} & 763.4 & 31.3 & 1.61$%
_{-0.20}^{+0.21}$ & 239$_{-16}^{+15}$ \\ 
\multicolumn{1}{l}{$14_{1}^{+}\rightarrow 12_{1}^{+}$} & 791.9 & 12.3 & 2.31$%
_{-0.35}^{+0.43}$ & 180$_{-17}^{+13}$ \\ 
\multicolumn{1}{l}{$16^{+}\rightarrow 14_{1}^{+}$} & 836.0 & 7.1 & 1.90$%
_{-0.27}^{+0.34}$ & 173$_{-16}^{+12}$ \\ \hline
\end{tabular}

\tablenotetext{1}{
Subscripts 1 and 2 indicate first and second excited state, respectively.} 
\tablenotetext{2}{
Uncertainties range from 3\% for strong to 40\% for the weakest transitions.}
\tablenotetext{3}{
Lower-limit value due to limited DSAM applicability.} 
\tablenotetext{4}{
Intensity-braching ratio of 0.26 $\pm $ 0.05 \cite{Snyder} was used (see text).}
\tablenotetext{5}{
$Qt$ value not shown in Fig. \ref{three}.} 
\tablenotetext{6}{
Intensity-braching ratio of 0.57 $\pm $ 0.13 \cite{Snyder} was used (see text).}
\tablenotetext{7}{
Transition reported in Ref. \cite{Wu-06}.} 
\end{table}%

\clearpage

\clearpage

\clearpage

\clearpage

\end{document}